%% file: main.tex
\documentclass[sigconf]{acmart}

\usepackage{tabularx}
\usepackage{colortbl}
\usepackage{multirow}
\usepackage{xspace}
\usepackage{amsmath}
\usepackage{fancybox}
\usepackage{subcaption}

\usepackage{amssymb}
\usepackage{booktabs}

\usepackage[strict]{changepage} %
\usepackage{framed} %

\usepackage{tcolorbox} %
\tcbset{colback=gray!2}

\newcommand{\threeStars}{\mbox{$^{***}$}}
\newcommand{\twoStars}{\mbox{$^{** }$}}
\newcommand{\oneStar}{\mbox{$^{*  }$}}

\newcommand{\grayrow}{\rowcolor{gray!10}}
\newcommand{\graycell}{\cellcolor{gray!10}}
\newcommand{\ra}[1]{\renewcommand{\arraystretch}{#1}}

\newcommand{\prReviewedBadge}{Pull Requests are Reviewed~\xspace}

\newcommand{\staticAnalysisBadge}{Static Analysis / Linters are Used~\xspace} 
\newcommand{\automatedTestsBadge}{Automated Tests are Run on Builds~\xspace} 
\newcommand{\unitTestsFastBadge}{Unit Tests are Fast~\xspace} 
\newcommand{\failedBuildsFixedQuicklyBadge}{Failed Builds are Fixed Quickly~\xspace} 
\newcommand{\deploymentAutomatedBadge}{Deployments are Automated~\xspace} 
\newcommand{\postDeploymentVerificationBadge}{Post-Deployment Verification is Automated~\xspace}  
\newcommand{\deploymentsOnCI}{Project has Automated Deployment CI Job~\xspace}
\newcommand{\trunkBasedDevelopment}{Project uses Trunk Based Development~\xspace}

\newcommand{\changeLeadTimeMetric}{Change Lead Time}
\newcommand{\cycleTimeMetric}{Cycle Time}
\newcommand{\timeToFirstCommitMetric}{Time to First Commit}
\newcommand{\meanTimeToResolutionMetric}{Mean Time to Resolution}
\newcommand{\averageReviewTimeMetric}{Average Review Time}
\newcommand{\ratioOfBugFixingCommitsMetric}{Ratio of Bug Fixing Commits}
\newcommand{\buildStabilityMetric}{Build Stability}
\newcommand{\normalizedCommitCountMetric}{Normalized Commit Count}
\newcommand{\normalizedPullRequestCountMetric}{Normalized Pull Request Count}
\newcommand{\normalizedReleaseCountMetric}{Normalized Release Count}

\newcommand{\rqone}{Is gamification effective as a means to promote the adoption of new practices?}
\newcommand{\rqtwo}{How does gamification impact the metrics of projects due to earning badges?}
\newcommand{\rqthree}{How do software developers react to gamification and perceive its impact?}

\definecolor{formalshade}{rgb}{0.95,0.95,1}
\definecolor{darkblue}{rgb}{0.0, 0.0, 0.55}
\newenvironment{quotebox}{%
    \vspace{-.1cm}
	\MakeFramed{\advance\hsize-\width\FrameRestore}%
	\noindent\hspace{-4.55pt}%
	\begin{adjustwidth}{}{7pt}%
		\vspace{-1pt}\vspace{1pt}%
	}
	{%
		\vspace{1pt}\end{adjustwidth}\endMakeFramed%
}

\newcommand{\conclusion}[1]{%
	\begin{tcolorbox}\vspace{-.1cm}#1\vspace{-.1cm}\end{tcolorbox}
}

\begin{document}

\title[Achievement Unlocked: A Case Study on Gamifying DevOps Practices in Industry]{Achievement Unlocked: A Case Study on Gamifying DevOps Practices in Industry}

\author{Patrick Ayoup}
\affiliation{%
\department{Department of Computer Science and Software Engineering}
\institution{Concordia University}
\city{Montreal}
\country{Canada}}
\email{p_ayoup@encs.concordia.ca}

\author{Diego Elias Costa}
\affiliation{%
\department{Department of Computer Science}
\institution{Université du Québec à Montréal}
\institution{LATECE Lab}
\city{Montreal}
\country{Canada}}
\email{costa.diego@uqam.ca}

\author{Emad Shihab}
\affiliation{%
\department{Department of Computer Science and Software Engineering}
\institution{Concordia University}
\city{Montreal}
\country{Canada}}
\email{eshihab@encs.concordia.ca}

\renewcommand{\shortauthors}{Ayoup, Costa, and Shihab}

\begin{abstract}
\input{abstract}

\end{abstract}



\maketitle

\section{Introduction}
\label{sec:introduction}
\input{introduction}

\section{Related Work}
\label{sec:relatedWorks}
\input{related_works}

\section{Context and Timeline}
\label{sec:background}
\label{sub:gamification-timeline}

\input{background}

\section{Case Study Design}
\label{sec:methodology}
\input{methodology}

\section{Results}
\label{sec:results}
\input{results}

\section{Discussion}
\label{sec:discussion}
\input{discussion}

\section{Threats to Validity}
\label{sec:threats}
\input{threats}

\section{Conclusion}
\label{sec:conclusion}
\input{conclusion}

\bibliographystyle{ACM-Reference-Format}
\bibliography{main}

\end{document}

%% file: abstract.tex
Gamification is the use of game elements such as points, leaderboards, and badges in a non-game context to encourage a desired behavior from individuals interacting with an environment. Recently, gamification has found its way into software engineering contexts as a means to promote certain activities to practitioners. Previous studies investigated the use of gamification to promote the adoption of a variety of tools and practices, however, these studies were either performed in an educational environment or in small to medium-sized teams of developers in the industry. 

We performed a large-scale mixed-methods study on the effects of badge-based gamification in promoting the adoption of DevOps practices in a very large company and evaluated how practice adoption is associated with changes in key delivery, quality, and throughput metrics of 333 software projects. 
We observed an accelerated adoption of some gamified DevOps practices by at least 60\%, with increased adoption rates up to 6x. 
We found mixed results when associating badge adoption and metric changes: 
teams that earned testing badges showed an increase in bug fixing commits but output fewer commits and pull requests; teams that earned code review and quality tooling badges exhibited faster delivery metrics. 
Finally, our empirical study was supplemented by a survey with 45 developers where 73\% of respondents found badges to be helpful for learning about and adopting new standardized practices. Our results contribute to the rich knowledge on gamification with a unique and important perspective from real industry practitioners.

%% file: introduction.tex
The tools, processes, and best practices in software development are constantly evolving as different trends emerge~\cite{Cico2021}. Although adopting new practices can be very attractive, provoking meaningful change and standardizing a heterogeneous environment at scale is a difficult task~\cite{Zulfahmi2019}. Getting developers, who have been using the same techniques for years, to change their ways is a challenge that needs careful thought and planning.

One creative solution to this problem is to incorporate gamification~\cite{Foucault2019}. 
Gamification is the inclusion of game elements in non-game context to motivate user activity and improve engagement~\cite{Deterding2011}.
Gamification has been reported to show positive results \cite{Singer2012,Prause2015,Foucault2019}, particularly when employed to promote the adoption of new tools and practices in software development~\cite{Singer2012,Dubois2013}. 
Although these studies showed promising results, the case studies were performed with students~\cite{Dubois2013,Khandelwal2017}, or a small to medium-sized teams of developers in industry~\cite{Garcia:20:GamificationSE,Neto2019,Foucault2019}.

Our paper complements the large body of work by performing a large-scale study of gamification and its impact in a real industrial environment. Specifically, we investigate the use of gamification over a year across 333 software development projects at a large company. 
In our case study, badges associated with DevOps best practices were presented to developers with the aim of improving certain key performance indicators (KPIs). 

We conduct a mixed-methods study to evaluate the relationship between gamification and the adoption of new practices. First, we study whether or not gamification is effective in promoting the adoption of new process and practices to see if it can act as an accelerant for changing behavior within an organization. Then, we investigate how the metrics of software development teams shift after making changes to their practices in order to earn badges. Finally, we surveyed practitioners working on these projects to learn how they react to gamification and perceive its impact. The aforementioned questions are of paramount importance to the studied organization (and others, we believe) to understand the effect of their efforts and how to better improve the existing gamification mechanisms.

Our work contributes to practitioners and the research community by: 
\begin{itemize}
    \item Presenting the first large-scale study on the effects of gamification on the adoption of DevOps practices in \textbf{industry}. Our study includes 333 projects from a large software development company. 
    
    \item Evaluating how changes in the DevOps practices encouraged by gamification are associated to changes in Delivery, Quality, and Throughput metrics of software projects. 
    
    \item Reporting qualitative insights from a survey with 45 industry practitioners about their reactions and perceived impact of gamification.  
\end{itemize}

This study provides a series of implications on gamification as a strategy to change practices in industry.
Our case study shows that gamification, if carefully designed, can be a powerful driver of new development practices, even in large and heterogenous industrial contexts. 
However, measuring the benefits of practice adoption using conventional delivery, quality, and throughput metrics can be difficult. 
Only some badges showed an association with project metrics change, and our results pointed to some trade-offs between quality metrics and development throughput.  
In addition, practitioners are driven by the benefits that gamified practices entail, and only to a lesser extent by the competitiveness and achievement provided by games.
For instance, practitioners were drawn to deployment and testing practices for the prospect of automation and reducing manual work and improving software quality.
Finally, we report on criticism and limitations that need to be addressed by the community to improve the effectiveness of gamification as a catalyst for behavioral change.

%% file: related_works.tex
In this section, we describe the fundamentals of gamification and dive in the related works that investigate gamification in SE. 

\subsection{Gamification and Motivation}

In its simplest definition, gamification is the application of game elements
and characteristics in a non-game environment \cite{Deterding2011}.
Gamification can manifest itself in
many forms by applying game elements such as points, badges, levels, quests, and
leader boards to support user engagement and enhance positive patterns~\cite{Hamari2014,Pedreira2014}.
Each game element has the potential to affect user behavior differently~\cite{Mekler:17:EffectGameElements}.
For instance, leaderboards emphasize relative performance and may drive users competitiveness~\cite{Mekler:17:EffectGameElements}, while badges, give the user a sense of self-improvement, and have shown to steer users' long-term behavior towards gamified goals~\cite{Hamari:17:Badges}.

Several studies have investigated the effects of gamification in a variety of domains~\cite{Seaborn:15:GamificationTheoryAction}. 
From education~\cite{Dicheva:15:Education} and health~\cite{Johnson:16:Health}, to marketing and commerce~\cite{Meder:18:GamficationEcommerce}, meta-studies have shown benefits of gamification on user engagement and satisfaction~\cite{Seaborn:15:GamificationTheoryAction,Johnson:16:Health,Hamari2014,Dicheva:15:Education}.
Still, studies have pointed out important limitations of gamification. 
Not all activities and contexts can be equally and effectively gamified. 
Users' perception of gamification vary considerably based on age and gender~\cite{Kovisto:14:PerceivedGamification}, user's receptivity to external rewards~\cite{Mekler:17:EffectGameElements}, and the meaning assigned to gamified elements~\cite{Cruz:17:PlayersPerception}. 
Finally, gamification's effectiveness is deeply connected to the design of game elements, and how they interact with the user~\cite{Mekler:17:EffectGameElements,Sailer:17:GamificationPerceived}.
A badly designed gamification system can sap user's motivation~\cite{Moldon:21:GamificationICSE,Toshihiko:13:BadGamification,Hanus:15:EffectsGamificationClassroom} and steer users to chase metrics instead of encourage behavioral change~\cite{Mekler:17:EffectGameElements}.

As a result, systematic studies unanimously state that more studies are needed to better understand gamification benefits and limitations~\cite{Nacke:17:MaturityGamification,Hamari2014,Dicheva:15:Education}. 
Particularly, researchers urge for large-scale studies that assess gamification effectivity on the long-term in the wild to complement studies in a lab environment~\cite{Nacke:17:MaturityGamification,Hamari2014}.    
Our study contributes to the literature by assessing gamification effectiveness in encouraging practitioners to adopt DevOps practices on a large software development company, over a period of one year.

\subsection{Gamification in Software Engineering}

Software engineering practitioners are no stranger to gamification. 
Major code-centric social platforms such as Stack Overflow use badges to evaluate users' commitment, competence and trustworthiness in the platform~\cite{Anderson:13:StackOverflow}.
Open-source projects in GitHub frequently employ badges to signalize to the community aspects related to the project quality, such as test coverage and build status~\cite{Trockman:18:BadgesOpenSource}.
Given its prominence, the effects of gamification has been studied in Software Engineering (SE) education~\cite{Alhammad:18:GamificationSEStudies}, and in open-source and industrial software development~\cite{Porto:2020:GamificationSE}.

Most works that study gamification in SE, focused on educational settings~\cite{Alhammad:18:GamificationSEStudies, Singer2012, Dubois2013, Prause2015, Khandelwal2017}.  
Alhammad and Moreno performed a systematic study on 21 papers that study gamification in SE education~\cite{Alhammad:18:GamificationSEStudies}. 
They found that gamification has reported mostly positive results in improving students engagement and, to a lesser extent, improving students knowledge. 
Dubois and Tamburrelli~\cite{Dubois2013} reported that students that participated in a gamified course showed better results, motivated by competition with their colleagues. 
Singer and Schneider, on the other hand, reported mixed results when employing gamification to encourage students to use control version systems more frequently~\cite{Singer2012}. 
Code review has also been gamified in a study by Kandelwal et al.~\cite{Khandelwal2017}. Comments originating from gamified systems were perceived as more useful by users, however the time needed to review the code was longer and uncovered a similar number of bugs in reviews from non-gamified environments.

Some works investigated gamification in open source software projects~\cite{Vasilescu2014, Moldon:21:GamificationICSE, Trockman:18:BadgesNPMEcosystem}.  
Vasilescu studied the engagement and contributions of developers to open source software projects and found that due to the recognition gamification provides, developers are more willing to engage in discussion and contribute more~\cite{Vasilescu2014}.
Open source software projects also commonly use badges to show to the community the adherence to good practices of software development (e.g., test coverage, build status), and Trockman et al. study showed that badges are mostly reliable as a signal of best practices~\cite{Trockman:18:BadgesNPMEcosystem}. 
However, gamification has also been shown to steer developers behavior towards unwanted directions~\cite{Moldon:21:GamificationICSE}, hence, the gamification system needs to be carefully designed.

Finally, a few studies have assessed gamification in industrial settings, most commonly in small and medium sized team of developers~\cite{Garcia2017,Neto2019,Foucault2019}.
Garcia et al. proposed a framework for incorporating gamification into software development tools and performed a case study at a small company with 19 practitioners~\cite{Garcia2017}.  
The authors reported seeing a 20\% increase in the usage of the requirement and issue tracking tools.
Neto et al.~\cite{Neto2019} developed a plugin for Redmine including several gamification elements and evaluated its effectiveness in a case study involving 19 developers from a small company. 
While many developers felt the gamification had positive effects on their work, the results were inconclusive as to whether or not developer productivity was improved.
Foucault et al.~\cite{Foucault2019} also performed an industrial case study with a system they built to gamify the adoption of good coding practices and the usage of static analysis tools involving 67 participants between two companies. Feedback from developers was mostly positive, showing that a sense of competition motivated developers to address static analysis warnings more seriously.

Our study complements above mentioned works by investigating gamification at scale in industry.
Over 300 projects which use a variety of technologies, and solve a number of different business problems are observed in this study. The developers building and maintaining these projects also have a wide range of professional experience levels and backgrounds. Additionally, this study looks at gamifying a variety of practices targeting different phases of the software development lifecycle, while past work mainly focused on one single aspect (i.e., version control).

%% file: background.tex
\begin{figure}
	\centering
	\includegraphics[scale=.6, trim=6cm 9.3cm 8cm 7cm,clip,page=2]{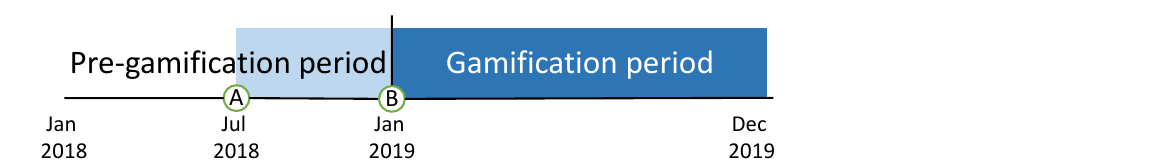}
	\caption{Timeline of the Gamification related events}
	\label{figures/gamification_timeline}
\end{figure}

This case study is centered around a large, multi-national company with a particularly large technology division comprised of more than 20,000 practitioners spread across the world. While development teams have the freedom to make decisions on the tools, technologies, and processes they adopt, there are a number of key best practices which should be more widely adopted. The gamification system under study is an initiative towards homogeneizing and promoting the best practices adopted by teams in the company.

In July 2018, an effort began to investigate DevOps best practices which would be beneficial for the development community. The output of this effort is a set of \textbf{DevOps Guidelines} suggesting which practices and tools a team is encouraged to prioritize and why they would be beneficial. These guidelines were socialized in July 2018 as marked by event A in Figure~\ref{figures/gamification_timeline}, and served as the basis for the badges in the studied gamification system. 
In December 2018, the gamification system was announced and detailed to the development community, and released for general use in January 2019 (event B). Given that each of these events build on each other on a path towards DevOps adoption, it is expected that the events leading to the deployment of the gamification system may influence the adoption of DevOps practices to some degree.
Hence, while we aim to study the effect of the gamification system (event B), we include event A in the study to control for eventual effects of the guideline in promoting the adoption of DevOps practices.

%% file: methodology.tex
The main vision behind the gamification system was to promote the adoption on DevOps practices with the ultimate goal of enabling software development teams to deliver more functionality, more quickly, while maintainining software quality and stability.
In this context, the design of our study centers on investigating the impact of gamification of DevOps practices under three main aspects: 

\textbf{RQ1: \rqone} We investigate how many projects worked towards earning the badges and what badges were more effective in encouraging the adoption of new practices. 

\textbf{RQ2: \rqtwo}
Naturally following from RQ1, for projects that are earning badges, this question investigates how each badge earned is associated with change in their key metrics.

\textbf{RQ3: \rqthree}
We conducted a survey with software developers to better understand their motivation to adopt badges and how they perceive their impact on their project metrics.

\begin{figure}[tb]
    \centering
    \includegraphics[width=.9\linewidth,trim={9cm 3cm 10cm 6.3cm}, clip]{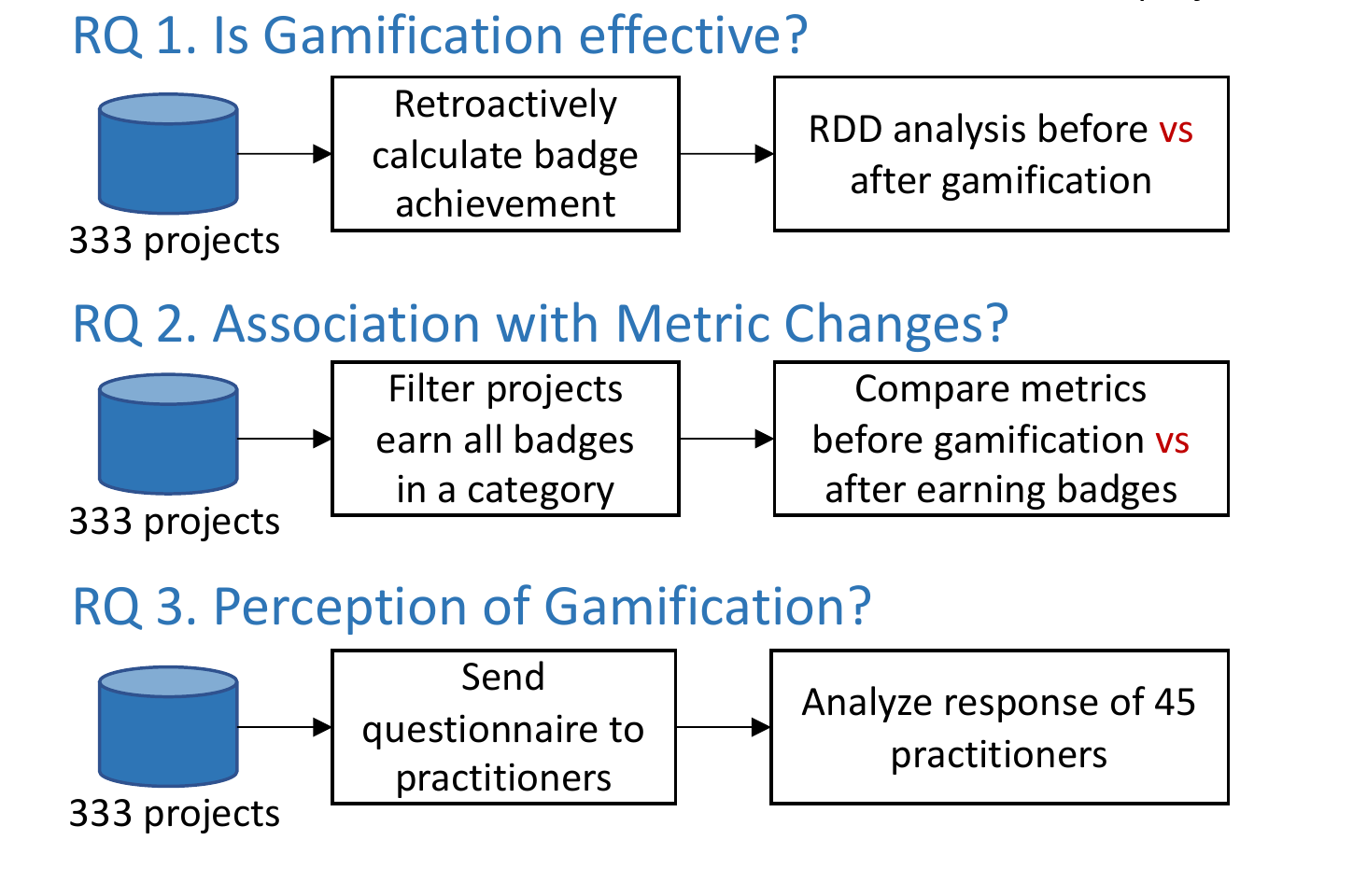}
    \caption{Methodology Overview}
    \label{fig:methodology-overview}
\end{figure}

Figure~\ref{fig:methodology-overview} presents a high-level abstraction of our methodology. 
In the remainder of the section, we describe the system of badges deployed by the company under study (Section~\ref{sub:badges}) and the metrics we select to evaluate the delivery, quality, and throughput of teams before and after gamification (Section~\ref{sub:metrics}).  

As this study was performed at a company in industry, the data used is proprietary and cannot be made available to the community. For this reason, numerical values are expressed in a relative form to properly retain anonymity.

\begin{table*}
	\caption{The badges considered in our study. The column "RQ1" depicts the badges that could be retroactively calculated for projects before the gamificaton period and are included in our RQ1 analysis. RQ2 and RQ3 analyses include all badges. } 
	\label{tables/badges}
	\small
	\input{tables/badges}
\end{table*}

\vspace{-0.1in}
\subsection{Badges}
\label{sub:badges}

The gamification system used in this case study leverages badges which are publicly displayed to the development community on each project's home page. For clarity of organization, badges are grouped into categories according to the following attributes of the software development lifecycle: deployment, git, quality tooling, review, stability, and testing.

A badge is a gamification element that serves as an indicator detailing whether or not the software development team working on a project has adopted a certain practice. Each badge is assigned an achievement requirement which must be met and maintained in order for it to be achieved. In order to encourage developers to adopt a given practice, a badge would be created targeting that practice with a requirement that can be evaluated to determine whether or not that practice has been adopted. For example, to encourage developers to adopt the practice of reviewing pull requests, a badge was created with the achievement requirement that at least 10\% of pull requests on a project must have evidence of review in order to achieve that badge.

The primary intention of deploying these badges is to encourage software development teams to learn new practices and to foster a sense of transparency and achievement. As such, all badges are awarded to a team working on a project, and not individuals. Additionally, it was one of the main design philosophies of the badges that they do not single out individual developers, or put teams to compete against eachother. For this reason, all achievement criteria for the badges are bound to properties of a project and not individuals (ie. number of PRs with evidence of review).

When the badges were announced to developers, it was explained what the badges were, and that their adoption was not mandatory in any way. Developers were informed that they simply serve to be recommendations of best practices and they are there to help if they wish to use them. Alongside the badges, documentation on how to achieve each one was made available to all developers. 
The badges considered in this study are outlined in Table \ref{tables/badges} along with the rationale for each badge's design.

\vspace{-0.1in}
\subsection{Metrics}
\label{sub:metrics}
The goal behind the implementation of gamification is to promote new practices that enable teams to deliver software more quickly while optimizing quality and stability. To assess this, we select metrics that cover different aspects of software delivery, quality and throughput. Delivery metrics allow us to evaluate how quickly a team is delivering new functionality, quality metrics give a signal as to how software quality changes with newly adopted practices, and throughput metrics give an image on the quantities produced at both the contributor level (commit and pull request counts) and product level as a team (release count). Each selected metric is described in Table \ref{tables/metrics}.

\begin{table*}[tbh]
	\centering
	\caption{The delivery, quality and throughput metrics considered in the study.}
	\label{tables/metrics}
	\small
	\input{tables/metrics}
\end{table*}

\vspace{-0.1in}
\subsection{Data selection}

The data used for this case study is extracted from the following three systems: JIRA, Git, and Jenkins.
A number of selection criterion have been chosen to filter the dataset down to a more homogeneus collection of mature software projects. 
In the following, we describe in detail the criteria used to select mature software projects which use JIRA, Git, and Jenkins consistently. 

We aim to evaluate the effects of gamification on teams that work on active and mature software development projects. To that aim, we start our filtering process by removing projects that are inactive, immature, or are personal projects. Active and mature projects are selected based on the criteria that they have regular activity in JIRA, Git, and Jenkins during 2018 and 2019 (the period of study), have had releases during these years, and are developed by a team of developers.
We also exclude monorepos and repositories composed of configuration files as the activities of the badges do not apply to these projects.

\begin{table}
    \small
	\caption{Descriptive statistics of the 333 selected projects.}
	\label{tables/selectedProjects}
	\input{tables/selectedProjects}
\end{table}

After our selection process, we identify 333 projects that are candidates for our study. 
As shown in Table \ref{tables/selectedProjects},  projects in our curated dataset have sufficiently long development time ($\sim5$ years), and are developed by large teams ($\sim30$ collaborators).

%% file: tables/badges.tex
\ra{1.1}
\begin{center}
    \begin{tabular}{ l l p{0.23\linewidth} p{0.23\linewidth} c}
        \toprule
        
        \textbf{Category} &
        
        \textbf{Badge} &
        \textbf{Requirement} &
        \textbf{Rationale} &
        \textbf{RQ1} \\
        
        \midrule

        \multirow{6}{*}{\textbf{Deployment}} &
        \deploymentAutomatedBadge &
        Automate deployment procedures &
        Save time with repetitive activities and avoid human error &
        \checkmark \\

        & 
        \graycell \postDeploymentVerificationBadge &
        \graycell Automate post-deployment verification procedures &
        \graycell Save time with repetitive activities and avoid human error & 
        \graycell \checkmark \\
                                 
        &
        \deploymentsOnCI &
        Project can be automatically deployed from CI Pipeline &
        Encourage teams to adopt continuous delivery & 
        \checkmark \\

        \midrule

        \multirow{3}{*}{\textbf{Git}} &
        \graycell \trunkBasedDevelopment &
        \graycell The majority of releases come from the same branch &
        \graycell Simplify development and release workflows. Promote the use of feature flags. & 
        \graycell \checkmark \\
        
        \midrule
        
        \multirow{2}{*}{\textbf{Quality Tooling}} &
        \staticAnalysisBadge &
        Run quality tooling as part of automated builds &
        Identify code smells earlier in the software lifecycle & 
         -- \\

        \midrule
        
        \multirow{2}{*}{\textbf{Review}} &
        \graycell \prReviewedBadge &
        \graycell At least 10\% of pull requests have comments by peers &
        \graycell Identify requirement and semantic errors earlier & 
        \graycell -- \\
        
        \midrule
        
        \multirow{2}{*}{\textbf{Stability}} &
        \failedBuildsFixedQuicklyBadge &
        Mean time to fix is under 24 hours &
        Keep target environment stable to enable continuous delivery & 
        -- \\
        
        \midrule
        
        \multirow{5}{*}{\textbf{Testing}} &
        \graycell \automatedTestsBadge &
        \graycell Run automated tests and persist test results for each build &
        \graycell Produce evidence of testing to improve confidence in more frequent changes & 
        \graycell \checkmark \\

        &
        \unitTestsFastBadge &
        Total unit test runtime is less than 5 minutes &
        Keep delivery pipeline flowing smoothly & 
        -- \\
        
        \bottomrule
    \end{tabular}
\end{center}

%% file: tables/metrics.tex
\ra{1.1}
\begin{tabularx}{\linewidth}{l l X X}
	\toprule

	\textbf{Category} &
	\textbf{Metric Name} &
	\textbf{Description} &
	\textbf{Rationale} \\

	\midrule
	
	\multirow{9}{*}{\textbf{Delivery}} &
	\changeLeadTimeMetric &
	Time elapsed from introducing a commit to its deployment in production \cite{Forsgren2019} &
	Quantifies the overhead of additional non-coding related activities \\
	
	&
	\graycell \cycleTimeMetric &
	\graycell Total time a JIRA issue is in an ``In Progress" state. &
	\graycell Quantifies the amount of development time spent on a JIRA issue. \cite{Power2015} \\
	
	&
	\timeToFirstCommitMetric &
	Time elapsed from the creation of JIRA issue to the first related commit &
	Quantifies the waiting period before the issue is first addressed \\
	
	&
	\graycell \meanTimeToResolutionMetric &
	\graycell Time elapsed from the creation of the JIRA issue to its resolution &
	\graycell Quantifies the total time an issue takes to be fully completed \\	
	
	&
	\averageReviewTimeMetric &
	The average time a pull request takes to be merged &
	Quantifies how much time is spend on review and reworking of pull requests. \\ 
	
	\midrule
	
	\multirow{5}{*}{\textbf{Quality}} &
	\ratioOfBugFixingCommitsMetric &
	Ratio of commits linked to fixing bug issues in JIRA vs all commits. &
	Quantifies how much work is targeted at fixing bugs vs delivering new features \\
	
	&
	\graycell \buildStabilityMetric &
	\graycell Ratio of successful vs unsuccesful builds in continuous integration, including compilation, automated tests and static analysis. &
	\graycell Indication of the overall health of the project. \\
	
	\midrule
	
	\multirow{6}{*}{\textbf{Throughput}} &
	\normalizedCommitCountMetric &
	Total number of commits normalized by the number of contributing developers &
	Quantifies the output of a team in terms of commits committed \\
	
	&
	\graycell \normalizedPullRequestCountMetric &
	\graycell Total number of pull requests merged normalized by the number of contributing developers &
	\graycell Quantifies the output of a team in terms of pull requests merged \\
	
	&
	\normalizedReleaseCountMetric &
	Total number of releases normalized by the number of contributing developers &
	Quantifies the output of a team in terms of releases for the client \\
	
	\bottomrule
\end{tabularx}

%% file: tables/selectedProjects.tex
\begin{center}
    \begin{tabular}{ l rrrr }
    	\toprule
		
		&
		\textbf{Median} &
		\textbf{Min} &
		\textbf{Max} \\

    	\midrule

    	\textbf{Project Age	(years)} &
		$\sim5$ &
		>3 &
		<20 \\

    	 \textbf{Total Commits} &
		 $\sim2000$ &
		 >100 &
		 <70,000 \\

    	\textbf{Authors} &
		$\sim30$ &
		>3 &
		<300 \\

    	\textbf{Files} &
		 $\sim1000$ &
		 >40 &
		 <20,000 \\
		
		\bottomrule
    \end{tabular}
\end{center}

%% file: results.tex
\subsection{RQ1. \rqone}
\input{rq1_results}

\subsection{RQ2. \rqtwo}
\input{rq2_results}

\subsection{RQ3. \rqthree}
\input{rq3_results}

%% file: rq1_results.tex
\noindent
\textbf{Motivation:}
A series of badges were designed and presented to users to encourage the adoption of new DevOps practices.
In this RQ, we investigate if these badges have helped promote the adoption of related DevOps practices and which badges had successful outcomes aiming to reach this goal.
While gamification has shown to be effective in many contexts~\cite{Hamari2014,Hanus:15:EffectsGamificationClassroom,Nacke:17:MaturityGamification}, we have yet to see its effectiveness on large software development companies.
Answering this question may shed the light on the benefits and limitations of gamification as a strategy for changing development practices.

\noindent
\textbf{Approach:}
Because each badge is associated with a practice, we evaluate whether gamification has helped accelerate adoption of the gamified practices. To investigate the effectiveness of badges in promoting new practices, we looked at the DevOps practices associated with each badge before and after gamification was implemented.
To that aim, we calculate the badge achievement status (whether or not a badge is earned by satisfying its requirement) retroactively for each month of the pre-gamification period. 
With the monthly badge achievement statuses in both periods, we compare the practice adoption in the pre-gamification period against the gamification period.

We compare the adoption of practices (badges) in both periods using data visualization and statistical modeling.
We employ the Regression Discontinuity Design (RDD)~\cite{Thistlethwaite1960}, an analysis that allow us to determine the longitudinal effects of an event on a time-series.
RDD is a quasi-experimental analysis that can be used to assess the discontinuity of a function as a result of an intervention, the gamification in our case. 
This method looks at the difference in a function's level and slope after an intervention with the assumption that without an intervention, the function would remain with the same level and slope.  
This method has been used in several previous studies to investigate the longitudinal impact of software engineering processes on software metrics~\cite{Zhao:2017:ImpactCISoftware,Trockman:18:BadgesNPMEcosystem,Zimmermann:19:SwitchingTrackers}. 

We use RDD to perform an analysis on the adoption of each badge individually where $Y$ is the total number of projects achieving that specific badge.
We specify the following linear regression model to estimate the level and slope in $Y$ before and after gamification:
\vspace{-.2cm}

\[Y = \alpha + \beta \cdot T + \gamma \cdot G + \delta \cdot A + \eta \cdot C + \epsilon_i\]

where $T$ represents \textbf{time} in months from the start of the observation period, 
$G$ is a binary flag indicating the period before \textbf{gamification} ($G= 0$) and after gamification began ($G  = 1$); 
and $A$ represents the number of months \textbf{after} gamification, coded 0 before gamification and incrementally increasing after gamification began.
In the \textbf{control} ($C$), we include the occurrence of Event~A (the DevOps Guidelines document described on Section~\ref{sub:gamification-timeline}), to control for effects caused by initiatives prior to the gamification.

This model is composed by two regressions.
Before gamification, the regression line has a $\beta$ + $\eta$ slope, and after gamification the slope changes to $\beta$ + $\eta$ + $\delta$.
The change in the regression level is the difference between the two regression values at the gamification starting point, and is given by $\gamma$.
We are interested in analyzing the change in the level ($\gamma$) and in the slope ($\delta$) of badge adoption once gamification is introduced.
For this analysis, we consider only those badges that were available at the inception of gamification and were related to practices we could reliably track and extract in both the pre-gamification and gamification periods. 
Hence, we only conduct the analysis on five of nine badges as shown in column "RQ1" in Table~\ref{tables/badges}.

\begin{table*}
	\caption{Adoption of the five badges achieved before gamification and the increase in adoption after a year of gamification.}
	\centering
	\small

\input{tables/popular-badges}
	\label{tables/popular-badges}
\end{table*}

\noindent

\begin{figure}[tbh]
	\centering
	
	\begin{subfigure}{0.49\linewidth}
		\centering
		\includegraphics[width=\linewidth]{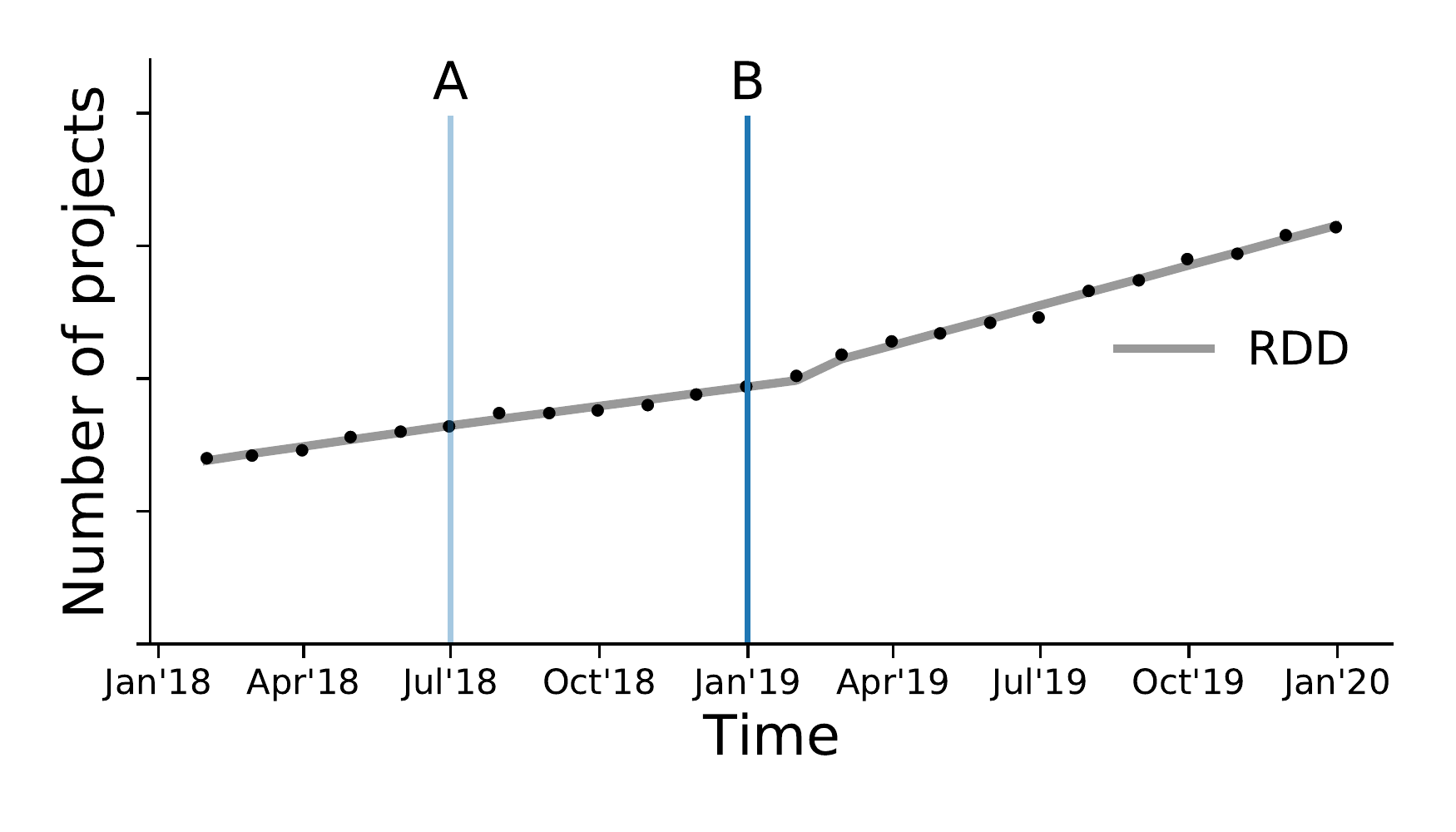}
		\caption{Deployments are automated.}
		\label{figures/rq1_Successful-automated-deployment}
	\end{subfigure}
~
	\begin{subfigure}{0.49\linewidth}
		\centering
		\includegraphics[width=\linewidth]{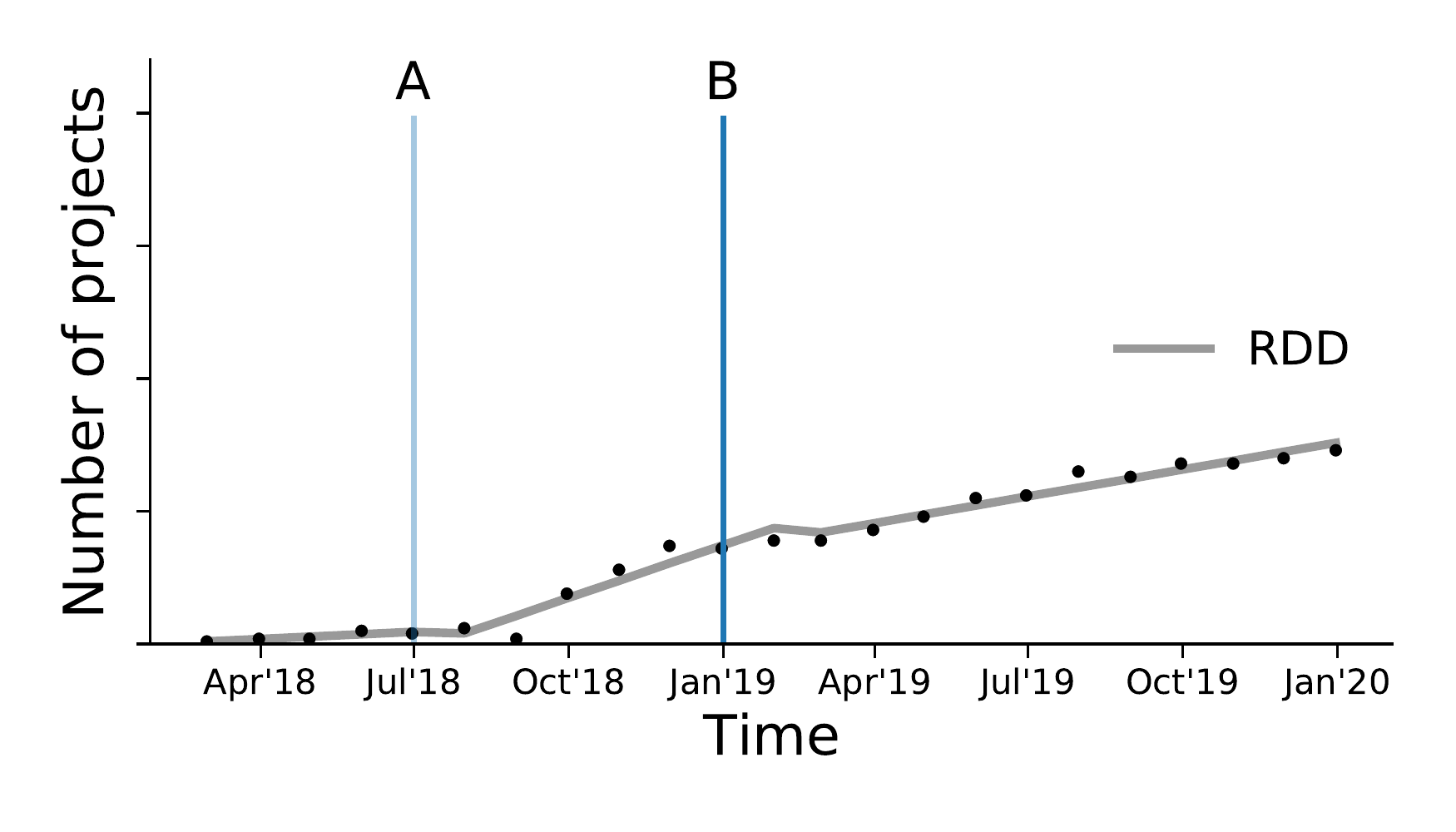}
		\caption{Project has deploy CI job.}
		\label{figures/rq1_Project-has-automated-deploy-job-configured}
	\end{subfigure}

	\begin{subfigure}{0.49\linewidth}
    	\centering
    	\includegraphics[width=\linewidth]{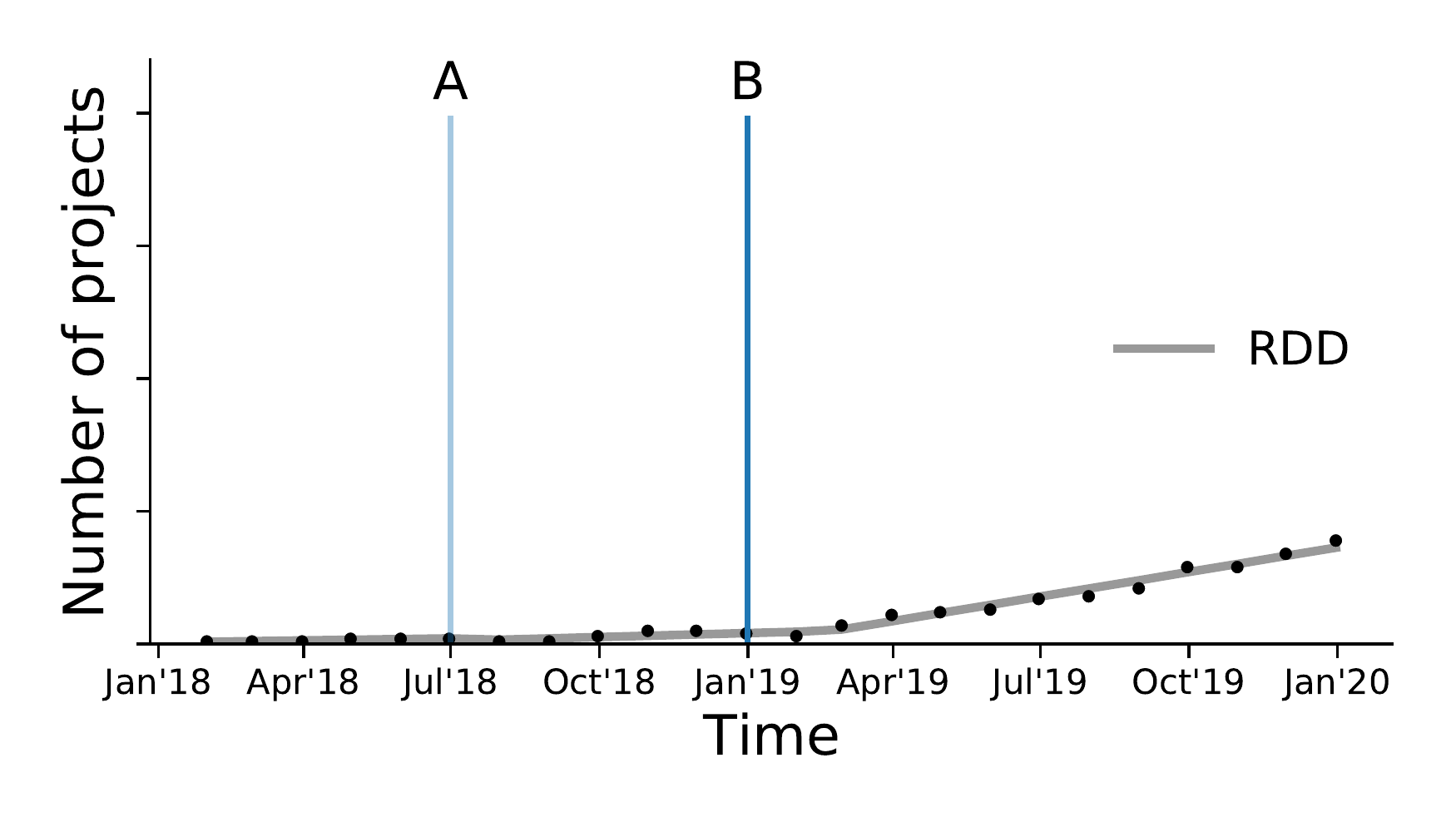}
    	\caption{Deployment Verification is Automated.}
    	\label{figures/rq1_Successful-verification-of-deployment-to-non-prod-environment}
    \end{subfigure}
~
	\begin{subfigure}{0.49\linewidth}
		\centering
		\includegraphics[width=\linewidth]{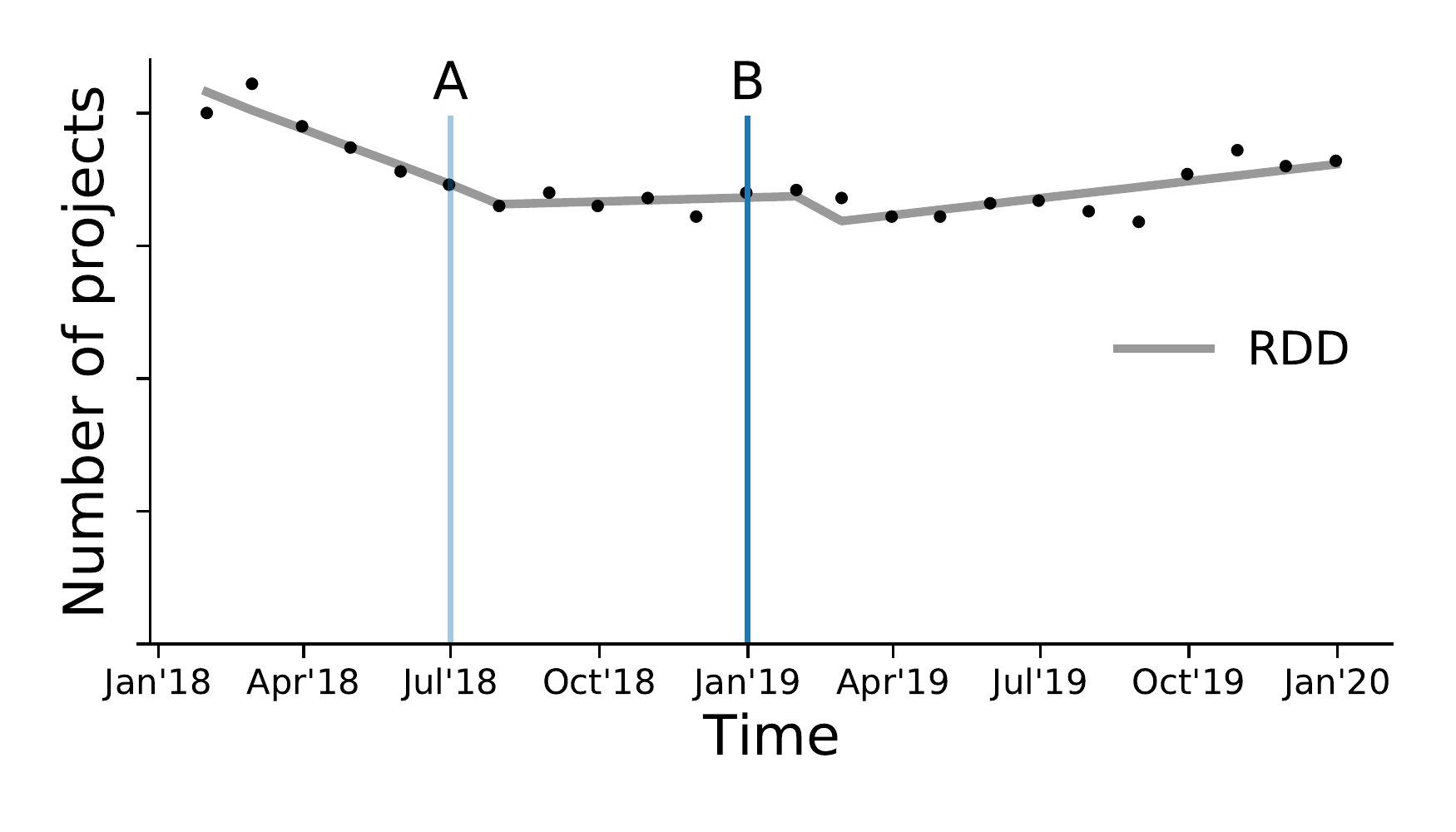}
		\caption{Project uses trunk based development.}
		\label{figures/rq1_One_branch_is_used_for_most_releases}
	\end{subfigure}

	\begin{subfigure}{0.49\linewidth}
		\centering
		\includegraphics[width=\linewidth]{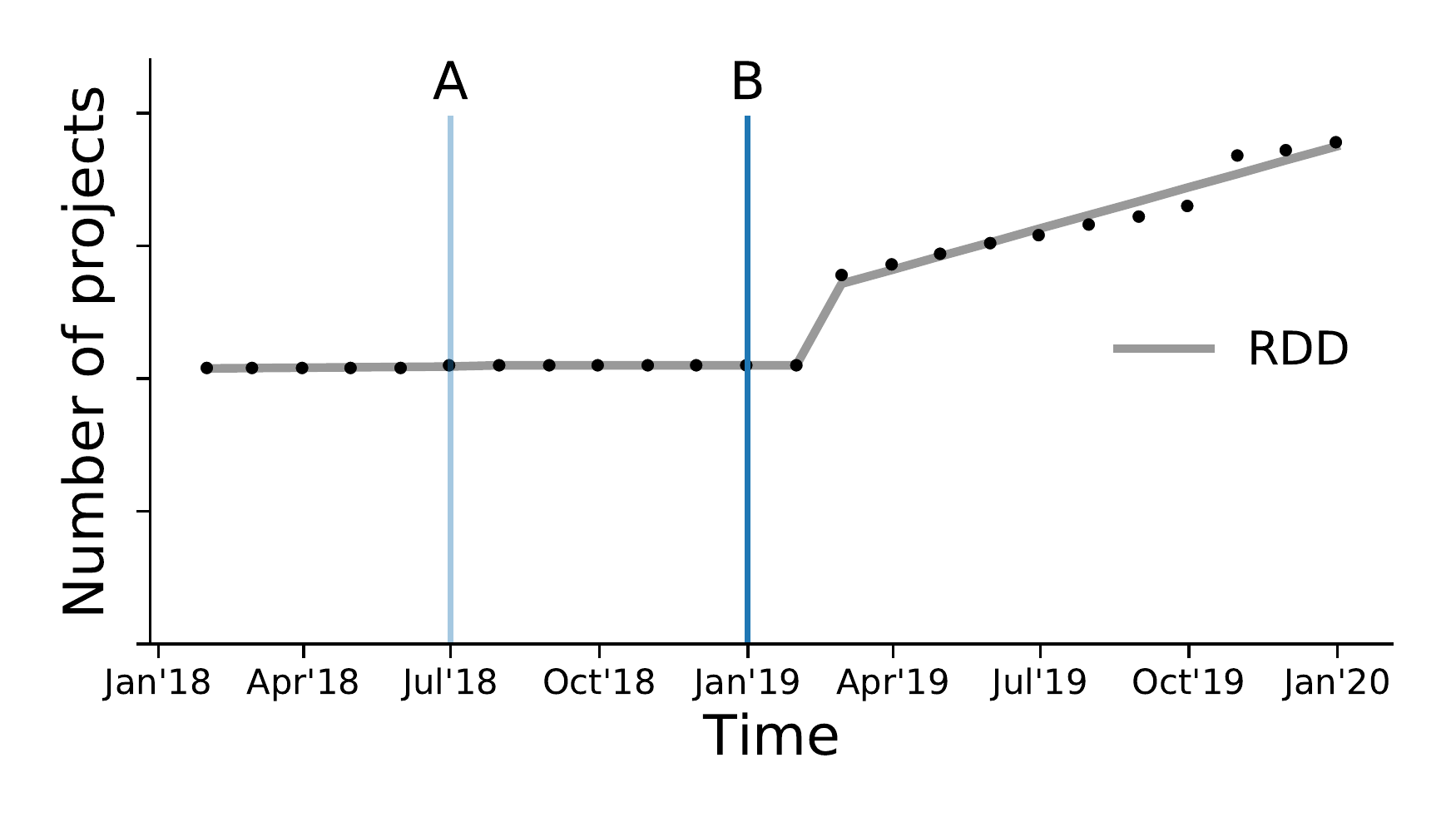}
		\caption{Automated Tests are Run on Builds.}
		\label{figures/rq1_Project-has-evidence-of-unit-tests}
	\end{subfigure}

	\caption{Evolution of DevOps practice adoption throughout the years of 2018 and 2019. 
	Event A refers to the release of the DevOps Guidelines and event B shows where Gamification began. 
	While we anonymize the y-axis values, we kept its proportion across badges to make the plots comparable. }
	\label{figures/badges-adoption-all}
\end{figure}

\noindent
\textbf{Results.}
To investigate which practices are seeing the most adoption, we analyze the adoption of each practice individually using the RDD analysis and the graphics shown in Figure~\ref{figures/badges-adoption-all}.
Table~\ref{tables/popular-badges} presents the adoption of the practices before gamification, the results of our RDD analysis, including the fitness of the model ($R^2$), the change in the level ($\gamma$), and the change in slope ($\delta$) caused by the introduction of gamification. Finally, we also present the improvement in the percent of projects adopting each practice by comparing the adoption level in the last month before gamification against the adoption level one year after the gamification.

\textit{\deploymentAutomatedBadge} (Figure~\ref{figures/rq1_Successful-automated-deployment}) initially had a moderate level of adoption and saw a significant increase in the slope of adoption ($\delta$ = 2.6) which corresponds to a 2x increase in the rate of adoption. \textit{\deploymentsOnCI} (Figure~\ref{figures/rq1_Project-has-automated-deploy-job-configured}) on the other hand did not experience a significant discontinuity due to gamification. This badge instead saw a short term steep increase in slope after the release of the DevOps guidelines, plateauing right before the gamification period began. While there was a slight increase of slope after gamification, it was not as large in magnitude compared to the effect of the DevOps guidelines. \textit{\postDeploymentVerificationBadge} (Figure~\ref{figures/rq1_Successful-verification-of-deployment-to-non-prod-environment}) saw the largest increase in adoption level (>650\%) and displayed a 6x increase in the rate of adoption ($\delta$ = 2.6). This practice in particular was newly implemented in the deployment tooling in the year before gamification, and thus had a very low initial adoption level. As seen in the analysis results, this practice benefited greatly from the education power of gamification. \textit{\trunkBasedDevelopment} (Figure~\ref{figures/rq1_One_branch_is_used_for_most_releases}) was the only practice with a high initial level of adoption, and also the only practice that did not see a statistically significant change in level or slope with gamification. Finally, \textit{\automatedTestsBadge} (Figure~\ref{figures/rq1_Project-has-evidence-of-unit-tests}) saw significant growth in adoption with an increase of >75\% from December 2018 to December 2019. While the other practices mainly saw a small rise in level post gamification and a significant increase in slope, this practice saw a substantial increase in level after gamification. This could be due to the fact that this practice is fairly well understood already and is an accessible starting point on the journey to adopting better practices.

\conclusion{We observed an accelerated adoption of gamified practices related to Testing and Deployment,  with increases in adoption rates from 60\% to 650\%. Gamification showed no significant influence on Git practices, widely adopted before gamification. 
}

%% file: tables/popular-badges.tex
\begin{center}
\begin{tabular}{lll|r rlrl|r}
	
    \toprule
        
    \multirow{3}{*}{\textbf{Category}} &
    \multirow{3}{*}{\textbf{Badges}} &
    \textbf{Adoption } &
    \multicolumn{5}{c|}{\textbf{RDD Analysis}} &
    \textbf{\% Improvement} \\
        
    &			
    &
    \textbf{Before} & 
    &
    \multicolumn{4}{c|}{} &
    \textbf{After Gamification} \\

    &			
    & 
    \textbf{Gamification} &
    \textbf{$R^2$} &
    \multicolumn{2}{c}{Level $\gamma$} &
    \multicolumn{2}{c|}{Slope $\delta$} &
    \textbf{(Dec. 2018 vs Dec. 2019)} \\

    \midrule

     \textbf{Git} &
     \trunkBasedDevelopment &
     High &
     0.84 &
     -11.5 &
     $\dagger$ &
     1.6 &
     $\dagger$ &
     $>$10\% \\
    
    \midrule

    \multirow{3}{*}{\textbf{Deployment}} &
    \deploymentAutomatedBadge &
    Moderate &
    0.99 &
    3.1 &
    $\dagger$ &
    2.6 &
    \threeStars &
    $>$60\% \\
        
    &
     \postDeploymentVerificationBadge &
     Low &
     0.98 &
     -2.2 &
     $\dagger$ &
     2.6 &
     \threeStars &
     $>$650\% \\

    & 
    \deploymentsOnCI &
    Low &
    0.98 &
    -5 &
    $\dagger$ &
    -3.2 &
    \threeStars &
    $>$95\% \\

	\midrule

    \textbf{Testing} & 
     \automatedTestsBadge &
     Moderate &
     0.99 &
     25.6 &
     \threeStars &
     5.2 &
     \threeStars &
     $>$75\% \\

    \bottomrule
\end{tabular}
\end{center}
\footnotesize
\begin{flushleft}
$\dagger p > 0.05$, \threeStars$p < 0.001$  \\
Low = adoption lower than than 20\%, Moderate = adoption between 20 and 60\%, High = adoption higher than 60\%
\end{flushleft}

%% file: rq2_results.tex
\noindent
\textbf{Motivation:} The DevOps badges were introduced with the primary aim of promoting the adoption of new DevOps practices and improving the overall software development process. While we observed in RQ1 the acceleration of the adoption of several gamified practices, in this RQ we examine if the adoption of these practices is associated with measurable changes of delivery, quality, and throughput metrics on the teams which adopt them.

\noindent
\textbf{Approach:} 
To examine whether there is an association between projects that earn badges and significant metric changes, we compare project metrics before and after badges are earned. 
Some badges, however, are complementary to each other as shown in the categories of badges of Table~\ref{tables/badges}. 
For example, \textit{\deploymentAutomatedBadge} and \textit{\postDeploymentVerificationBadge} are both concerned with the deployment automation process. Hence, it stands to reason that both badges are complementary in promoting a change in the deployment practices which may influence evaluated metrics.

To address potentially confounding effects from closely related badges, 
we evaluate the observed effect of earning all badges within a category on each of the selected metrics. 
For each badge category, we find the projects that have earned all of the badges in that category during the same month (e.g., all deployment badges). 
For each project in this subset, we calculate the mean value of each of our selected metrics over the last six months of the pre-gamification period (July - December 2018), and the first six months after that project earned the badges in the category under study (specific for each project). The result of this process is two distributions, one containing the mean values of a metric pre-gamification and one containing the mean values of a metric post-achievement. These two distributions are then tested for significant changes in each of our selected metrics using the Wilcoxon Signed-Rank Test~\cite{Wilcoxon1945}.
To quantify the effect size of statistically significant changes, we resort to the Cliff's Delta effect size~\cite{cliff:1993:dominancestatistics} and use Romano et al~\cite{Romano} guide for interpreting the effect size, similarly to previous works~\cite{Costa:19:JMH,Wessel:18:Bots}.

\begin{table}[tb]
	\caption{Relationship between earning badges and  metrics. 
	We only present the 7 out of 60 evaluated combinations (badge category x metrics)  which showed statistical significant differences.}
	\centering
	\small

\input{tables/rq2_results}

	\label{tables/rq2_results}	
\end{table}

\noindent
\textbf{Results:} 
We evaluate how badges from 6 different categories (deployment, git, quality tooling, review, stability, testing) affect the 10 metrics related to the aspects of delivery, quality and throughput.  
Hence, we evaluate 60 combinations total (6 badge categories x 10 metrics). 
After evaluating each of these combinations, seven of these combinations showed a statistically significant change in the mean value after earning the associated badges. Table~\ref{tables/rq2_results} summarizes the associated impact of the various badge categories on metrics for the cases where significant change has been observed. Cases where no significant change has been observed are omitted from this table for the sake of brevity. 

The results of this analysis show badges related to review, quality tooling, deployment, and testing were overall associated with a small to moderate effect on the selected metrics. We have observed both positive effects suggesting that teams that earn badges had an associated improvement in some metrics while also seeing negative effects for other metrics. 
These results outline a possible tradeoff which are paid when earning the associated badges.

\noindent
\textbf{Delivery Metrics:}
When considering how earning badges are associated with changes in the delivery metrics of a team, we observed that the most impactful badges are the review and testing related badges. 
Teams that earn review badges had exhibited an improvement in \textit{\changeLeadTimeMetric}~(negative Cliff's delta), indicating that individual commits are reaching production faster after earning the badges. 
However, teams that earned the testing badges showed a slow down of the overall resolution time of JIRA issues (\textit{\meanTimeToResolutionMetric}), evidenced by the positive Cliff's delta of medium magnitude.
Secondary to these badges, teams that earned quality tooling badges exhibited a slightly longer \textit{\cycleTimeMetric}~to finish their JIRA issues (positive Cliff's delta). 
This suggests that using code quality tooling may be associated with reducing the total development time alotted to a given JIRA issue.

\noindent
\textbf{Quality Metrics:}
Of the metrics studied, only \textit{\ratioOfBugFixingCommitsMetric}~showed any significant change after teams acquired any of the studied badges, i.e., we notice no significant change in the \textit{\buildStabilityMetric}~(our complementary quality metric).
Teams that earn testing badges had shown a positive change in \textit{\ratioOfBugFixingCommitsMetric}, of medium effect size (positive Cliff's Delta). 
This suggests that after achieving the testing badges, software teams have observed a larger proportion of commits are linked with bug issues in JIRA compared to before the gamification period.

\noindent
\textbf{Throughput Metrics:}
Overall, the only categories in which we identify a significant change of throughput metrics after teams acquire the badges has seen only negative effects.
Teams who earned the testing badges saw a negative effect for both \textit{\normalizedCommitCountMetric}~ and \textit{\normalizedPullRequestCountMetric}, suggesting that they are producing fewer commits and pull requests than before gamification. 
Additionally, projects earning the deployment badges saw a medium sized negative effect on the \textit{\normalizedPullRequestCountMetric}~ metric, indicating that these teams are outputting fewer pull requests after earning the badge. 

\conclusion{
	We found significant changes in 7 of the 60 metric / badge category combinations.
	Teams that earned Testing badges showed an increase in the number of bug fixing commits, but output fewer commit and pull requests. 
	Teams that earned Code Review and Quality Tooling badges have exhibited shorter change lead time and cycle time metrics.}

%% file: tables/rq2_results.tex
\begin{center}
\begin{tabular}{llr@{\hspace{0.5\tabcolsep}}lr}
\toprule

\textbf{Category} &
\textbf{Metric} &
\multicolumn{2}{c}{\textbf{Cliffs Delta}} &
\textbf{Proj.} \\

\midrule

\multirow{1}{*}{\textbf{Deployment}} &

\textbf{\normalizedPullRequestCountMetric} &
-0.400 \oneStar &
M &
10 \\

\midrule

\textbf{Quality} &

\multirow{2}{*}{\textbf{\cycleTimeMetric}} &
\multirow{2}{*}{-0.322 \oneStar} &
\multirow{2}{*}{S} &
\multirow{2}{*}{17} \\

\textbf{Tooling} &
 &
 &
 &
 \\

\midrule

\multirow{1}{*}{\textbf{Review}} &

\textbf{\changeLeadTimeMetric} &
-0.357 \twoStars &
M &
19 \\

\midrule

\multirow{4}{*}{\textbf{Testing}} &

\textbf{\meanTimeToResolutionMetric} &
0.385 \oneStar &
M &
27 \\

&
\textbf{\ratioOfBugFixingCommitsMetric} &
0.384 \twoStars &
M &
27 \\

&
\textbf{\normalizedCommitCountMetric} &
-0.276 \twoStars &
S &
27 \\

&
\textbf{\normalizedPullRequestCountMetric} &
-0.267 \twoStars &
S &
27 \\

\bottomrule
\end{tabular}
\end{center}
\footnotesize
\begin{flushleft}
$^{**}p < 0.01$, $^{*}p < 0.05$ on Wilcoxon Signed Rank test. 
\end{flushleft}

%% file: rq3_results.tex
\noindent
\textbf{Motivation:}
Badges are expected to increase the adoption of certain processes and invoke change on a team's key metrics, both which are effects that can be measured directly. 
However, at its core, badges are designed to invoke change in developer's behavior. 
Hence, it is important to get quality feedback from developers adopting these practices to understand 1) how they feel about the badges and 2) to get a sense on any unmeasurable outcome the gamification may have in our study case.

\noindent
\textbf{Approach:}
In this RQ, we design and distribute a survey invitation to 600 developers who have contributed to the projects under study and have worked in the company through the inception of gamification on their projects. In order to avoid biased answers and encourage participants to answer truthfully, participants were informed that this was an anonymous survey when invited to participate. We received a total of 45 responses from the invited participants, resulting in a 7.5\% response rate, similar to the response rates in other surveys in software engineering research~\cite{Hoyos2021}.

Our survey is composed of two sections.
In the first section, we ask for background information about the respondent such as their role, the amount of experience they have, and the size of their team. In the second section, we ask a series of open-ended questions about the participant's perception towards gamification, their motivation for adopting or not adopting badges, and the perceived outcomes on their projects. To detect recurring themes in these responses, the first two authors independently classified them using an open card-sort method \cite{Fincher:05:OpenCardSort}.
Labels were created while evaluating the responses and new labels were retroactively applied wherever applicable.
The annotators then met to discuss their labeling and reach a consensus.
This process enabled us to observe which themes are most common across all survey respondents.

\begin{table}[tbh!]
	\caption{Results of biographical survey questions.}
	\centering
	\small
	\input{tables/rq3_biographical_info}
	\label{tables/rq3_biographical_info}	
\end{table}

\noindent
\textbf{Respondent Demographics:} 
We detail the demographical information of survey participants in Table~\ref{tables/rq3_biographical_info}.
Our participants cover a variety of roles in the company, with the majority being developers (26) and tech leads (11). 
Almost half of the respondents (22) have more than five years of experience in their respective areas, while an additional 19 respondents have 2-5 years of experience. The majority of our respondents are in medium to large sized teams.

\subsubsection*{\large What \textbf{motivates} you to achieve DevOps badges?\\}

The intent of this question is to uncover what motivates the survey respondents to use the badges and adopt their associated practices. 
The developers surveyed had a wide range of motivations for adopting badges, from the boosted automation of deployment practices to friendly competitive environment.

\textbf{Reduce manual overhead in their deployment process (18 respondents)}.
The most commonly cited motivating factor for adopting DevOps badges was that developers saw them as a guide to adopting new practices with the hope of reducing manual overhead. In their responses, survey participants detailed that they would like to reduce overhead primarily in the deployment process, but also in their testing processes. Additionally, with the reduction of manual overhead, they also suggest motivation by a reduction in manual error as a result of less manual intervention in these repetitive tasks.

\begin{quotebox}
	\textit{``Ease of code development and deployment process. Also, the fact that the deployments can be done with little to no risk. It also takes out any dependency from deployment and development team members'' - R14}
\end{quotebox}

\textbf{Adopt standardized tooling and processes over custom solutions (11 respondents)}.
Eleven respondents specified that they are encouraged to achieve the badges because they make it clear what is the standard tooling to adopt across their projects so that skills learned are reusable and transferrable between projects. 
This suggests that there is a drive to make these new processes repeatable and more easily supported by adopting tooling and processes which are standardized througout their environment.
\begin{quotebox}
	\textit{``Standards that enforce every project to do the same'' - R9}
\end{quotebox}

Other notable motivations include enjoying a sense of accomplishment from seeing progress and friendly competition with colleagues (5 respondents), and a motivation to earn badges as a means to showcase their achievements to others (2 respondents). Interestingly, only 3 respondents stated their motivation came from a top-down incentive from management, and 2 other participants were motivated to earn badges because they were helpful in justifying the improvement of internal processes to management. This is a particularly encouraging result as it suggests that badges are helpful for encouraging teams to be self-starters and take initiative to make change rather than being asked by their superiors.

\conclusion{Participants are driven to achieve DevOps badges as they showed a pathway to reducing manual overhead, and standardizing process across teams. Participants also enjoyed seeing accomplishment in adopting practices, and friendly competition with their colleagues.}

\subsubsection*{\large Are badges helpful in \textbf{adopting} DevOps practices?\\}

We designed a two-part question to investigate 1) if practitioners found badges helpful for guiding them to try and adopt new practices and 2) an open-ended question to elaborate on why (and why not) badges were deemed helpful.

Overall, 73.3\% of the survey respondents answered "yes" when asked whether or not they found badges helpful. 
When elaborating on why they found badges helpful, we received reponses which apply to the badges in general. These themes are as follows:

\textbf{Badges are useful for informing developers about better practices (8 respondents)}.
The most popular theme reported is that developers appreciated how badges provide a clear outline of what they should be adopting as best practices and what they should do to adopt them. The educational power of the badges can be very strong. One clear example of this is \textit{\postDeploymentVerificationBadge}. Reviewing Table \ref{tables/popular-badges} from RQ1, we can see that after the badge was created, the adoption level grew from very low by a large margin. Furthermore, in their elaboration, 7 respondents explained that the deployment badges were helpful to teach them about automated deployment practices. This feedback from users similarly supports the findings from RQ1 indicating that the associated badges (\textit{Deployments are Automated}, and \textit{Post-Deployment Verification is Automated}) were effective in helping a significant number of teams adopt new practices related to their deployment processes.

\begin{quotebox}
	\textit{``The associated posts which describe the badges, why it is a recommended practice and how to achieve it are invaluable tools for teams that are onboarding''} - R39
\end{quotebox}

\noindent
\textbf{Badges help improve transparency and communication (7 respondents)}.
While the main intent of badges are to promote the adoption of new practices, survey respondents noted that they are also quite helpful as a dashboard to provide transparency into the status of projects in terms of hygiene of their practices. Having this global view on their project is helpful to determine which practices they should be adopting.

\begin{quotebox}
    \textit{``The badges have helped us identify at a repo and more macro levels where we need to invest devops effort.''} - R39
\end{quotebox}

Participants also mentioned that badges were helpful for 
standardizing tooling and processes across teams (5 respondents), and setting concrete targets for improving current DevOps practices (2 respondents).

As for the 26.7\% of respondents who answered that they did not find the badges helpful, criticisms which were stated suggest that the badges takes a lot of effort for maintaining a positive appearance to peers, and this may drive the wrong motivations for teams to change their behavior. 
As a respondent stated: 
\begin{quotebox}
\textit{"I'd like to highlight that some may prioritize DevOps achievement in a wrong way 
which is steering away the focus on the actual KPI 
- this is a big problem as people are just getting badges for the sake of getting it to show off instead on worrying on the actual outcomes".} - R11
\end{quotebox}
 
Additionally, some respondents mentioned they had already adopted other practices which were working for them but contradict what the badges promote. This suggests a frustration that their previous efforts may be wasted or not recognised, and they felt a pressure to change their practices:

\begin{quotebox}
	 \textit{"We had already adopted most of the best practice that the badges are trying to make us adopt. Being forced to try and keep the metrics right is costing us time and forcing us to change our already established practices that were working well"} - R6.
\end{quotebox}

\conclusion{73\% of participants find badges helpful explaining that they inform teams about better practices and improve transparency and communication. Approximately 27\% of respondents did not find badges helpful, stating it may drive the wrong motivation for teams to change behavior.}

\subsubsection*{\large Did you perceive tangible \textbf{benefits} of adopting DevOps badges?\\}

The intent of this question is to examine the perceived results by the survey respondents on their projects as a result of adopting the practices associated with the DevOps badges. Of the surveyed participants, 62.2\% of respondents noted they observed benefits after earning badges, citing the following reasons: 

\noindent
\textbf{Reduction of manual overhead in deployment processes and an increase in deployment frequency (13 respondents)}.
The most frequent answer from respondents suggest that automated deployment practices have significantly reduced the complexity, overhead, and stress of deployments and improved quality of life for practitioners, ultimately improving their deployment frequency metrics. 
Also, there were reports of attitudes towards change management shifting as the badges which promote automated deployment enable more frequent deployments. One respondent reported that their team feels more secure with automation in place and this has changed their outlook on change management.
\begin{quotebox}
	 \textit{``Smoother deployments, more frequent deployments, easier to release many projects (no difference between releasing 1 project or 20 projects)''} - R16
\end{quotebox}

\noindent
\textbf{Improve testing practices and software quality (10 respondents).}
Aside from gains derived from automating deployments, developers also noted that they observed earning testing badges had positive outcomes on software quality. Specific outcomes quoted include repayment of technical debt, an increase in unit test coverage, and a perceived increase in software quality.
\begin{quotebox}
	\textit{``Introduced code quality tooling that helped with test coverage and technical debt. Gave up some bad practices of merging PRs without review.''} - R20
\end{quotebox}

Other tangible benefits mentioned by participants were the standardization of tooling and processes (3 respondents) and improving transparency of processes and communication in a team (3 respondents).
Participants mention, once again, that the badges have helped convince management to improve internal processes (2 respondents).

From the survey participants, 37.8\% reported not identifying tangible benefits from adopting badges. Only one of these participants elaborated on this by stating it was too early for them to tell whether or not there are any tangible benefits. Participants also provided other valuable feedback. 
One respondent mentioned that changing their practices negatively impacted their productivity because their current practices were already working well for their team. 
Similarly, respondents suggested that changing behaviors made their developers nervous about doing things which would cause them to lose a badge, 
an unintended consequence of gamification. For example, given the badge \unitTestsFastBadge, which requires tests to run in less than 5 minutes, a participant stated:

\begin{quotebox}
	 \textit{"Some tests take time to run, how do we make sure we run all the tests and [the] metric does not get affected?"} - P13
\end{quotebox}
 
\conclusion{The majority of participants (62\%) reported perceived tangible benefits of adopting DevOps badges. Gamified DevOps practices have reduced manual overhead in deployment and improved software quality and test practices. Still, 38\% report not identifying tangible benefits, with some complains of lower productivity and unintended consequences.}

%% file: tables/rq3_biographical_info.tex
\begin{center}
    \begin{tabular}{ lr|lr|lr }
    	\toprule
    	\textbf{Role}             
    	& \textbf{\#} 
    	& \textbf{Experience}   
    	& \textbf{\#}
    	& \textbf{Team Size}       
    	& \textbf{\#}
    \\
    	\midrule
    	
    	Developer                            
    	& 26 
    	& $<$ 2 Years  
    	& 4  
    	& 1-3 Members
    	& 1  
        \\
    	
    	\grayrow
    	Tech Lead                            
    	& 12 
    	& 2-5 Years    
    	& 19 
    	& 4-5 Members    
    	& 12 
        \\
    	
    	Infrastructure 
    	& \multirow{2}{*}{3}  
    	& \multirow{2}{*}{6-10 Years}   
    	& \multirow{2}{*}{8}  
    	& \multirow{2}{*}{6-10 Members}   
    	& \multirow{2}{*}{19}
    	\\
    	
    	Op. Engineer 
    	& 
    	& 
    	& 
    	& 
    	& 
    	\\
    	
    	\grayrow
    	Architect                            
    	& 2  
    	& 11-15 Years  
        & 8  
        & 11-15 Members  
        & 5  
        \\
        
    	QA                                   
    	& 1  
    	& 16-20 Years  
    	& 2  
    	& 16-20 Members
    	& 2  
    	\\
    	
    	\grayrow
		Product Owner                        
		& 1  
		& $>$ 20 Years 
		& 4  
		& $>$ 20 Members 
		& 6  
        \\

        \bottomrule
    \end{tabular}
\end{center}

%% file: discussion.tex
In this section, we discuss four overarching themes that emerge from the findings of our three RQs, which serve as implications to practitioners and researchers on the effectiveness of gamification. 

\textbf{Deployment and testing practices are good candidates for effective gamification.}
The results of our study indicate that deployment and test practices exhibited the best outcome of the gamified practices in the company under study. 
Of the badges we evaluated in this study, testing and deployment badges have shown to yield the highest growth in adoption following the implementation of gamification (RQ1). 
Teams that earn testing badges are associated with an increase in the number of bug fixing commits (RQ2). 
Related studies have also reported that gamified testing has yielded improvements in defect registration~\cite{Porto:2020:GamificationSE}.
Finally, practitioners frequently cite the reduction of manual overhead in deployment processes as the main motivation for using the badges (RQ3), and report perceived improvement in software quality and testing practics (RQ3). 
Our findings are also corroborated by related work, which cites Product Integration (Deployment) and Verification and Validation (Testing) as most frequently cited areas in which gamification exhibited a positive outcome~\cite{Porto:2020:GamificationSE}.

\textbf{Not all badges show an association with project metrics change.}
When considering how metrics change with the implementation of gamification, in RQ2 we saw a small fraction of badge category / metric combinations showing any significant change after teams aquired the relevant badges. Not all of these observed associations are positive. While an association between adopting the testing practices and a higher bug fixing commit ratio was seen, testing badges were also associated with a reduction in throughput metrics. As such, when encouraging new practices, there may be trade-offs between the KPIs associated to the practices adopted by a team. It is important to note that analysing changes on a large heterogeneous set of projects is a complex task, and confounding factors could interplay. Additionally, these associations observed do not suggest causation.

\textbf{Benefits of badges are not easily measurable.} 
Interestingly, when comparing the results from RQ2 with the survey responses in RQ3, we observed a contrast between developer perception and the measured metrics. While many participants have mentioned the deployment badges to be a game changer, we did not observe any positive outcomes in the evaluated metrics. Whether or not the badges produce concrete change in the evaluated metrics, they may impact the perception of developers on their processes and improve their quality of life in their work. In the future, more studies should be conducted to establish and/or confirm a link between the practitioners' perception and the result in their KPIs.

\textbf{Gamification systems need to be carefully designed.}
Although the survey participants had a lot of positive feedback about gamification, there were also a number of critics. One survey respondent expressed that they fear gamification can potentially drive the wrong motivations for change. Gamification may drive some developers to adopt the badges solely to check off all of the boxes and show off without being mindful of the underlying KPIs which are meant to be optimized by the badges. Another respondant also expressed fear that developers will waste too much time to maintain their badges, even if they are not actually deriving any real benefit, simply for the sake of vanity. In their study, Porto et al~\cite{Porto:2020:GamificationSE} also noticed the same problem in four of the studies they reviewed~\cite{Scherr2018,Johansson2014,Dalpiaz2017,DeMelo2014}, it is difficult to manage motivations and get users to focus on the right things.

\vspace{-.1cm}

%% file: threats.tex
In this section, we discuss the threats to the validity of our findings, broken down by internal, construct, and external validity.

\noindent
\textbf{Internal Validity.}
Threats to internal validity are related to experimenter bias and errors. 
First, analysing data from a large set of projects from a real world enterprise in a heterogenous environment was very challenging and errors in this process could affect our results. 
We mitigate this risk by including only the badges related to practices we could reliably track and extract from studied projects, leading us to remove four badges in our analysis in RQ1. 
Second, many factors could influence software developers to adopt DevOps practices, other than gamification, such as seeing examples of positive outcomes from other teams and companies. 
For this reason, we detailed event A in Figure~\ref{figures/gamification_timeline} to represent the communication of the DevOps Guidelines document. 
In RQ1, this is factored in as a control variable to observe how strongly it impacts our results. 
Similarly, in RQ2, practices which target the same KPI (ie. deployment related practices) may have confounding associated effects. In order to address this, we focused on the effects of groups of practices and observed how the targeted KPIs change. 
Even with our mitigation, we were careful to describe the metrics change as an association (not causation) with gamification, as there could be many other reasons explaining the change of a KPI metric. 
Finally, in RQ3, surveys can be subject to human error and bias. 
We mitigate this risk by submitting our survey to a large sample group from different areas of the company in attempt to get a full viewpoint of how individuals in different working situations view gamification.

\noindent
\textbf{Construct Validity.}
Our study uses a number of metrics to help assess the changes projects go through after practices are adopted and badges are earned. 
Some of these metrics, however, attempt to measure constructs of a software project which are not easy to measure (e.g., software quality). 
For instance, the quality metric \textit{\ratioOfBugFixingCommitsMetric} can be viewed in two contradictory manners. 
Having a high ratio of bug fixing commits can be viewed as a project having a lot of bugs, but it can also be viewed as a team being very active in improving the quality of their system. 
For this reason, the power of this metric in isolation is relatively low and it is best used in combination with other metrics to help better explain the state of a project. 
As for the throughput metrics, these metrics in isolation do not give the full picture of productivity as teams can have a variety of habits delivering functionality. 
More frequent releases could be more desirable, however, it is not safe to generalize that a team with this practice is more productive than a team which does larger, less frequent releases. 
The release size metric could help complement our analysis, but the data was unavailable for this study.

\noindent
\textbf{External Validity.}
This study took place in a large company with a distinct software development culture and approach. 
Other companies may not operate in the same way, and therefore the findings of this study may not be generalizable to all companies.

%% file: conclusion.tex
In this paper, we conducted a mixed-methods study on the effects of badge-based gamification at a large company. 
We investigated how badges can accelerate the adoption of new practices and their associations with a set of key delivery, quality, and throughput metrics. 
We also conducted a survey with practitioners to understand how developers react to gamification and perceive its impact. 
Our findings showed that gamification can be effective in promoting the adoption of new practices, with practice adoption increasing at least 60\% in most practices. 
Teams that earned badges related to code review and code quality tooling saw a small to moderate reduction in their cycle time and change lead time metrics. 
Additionally, teams which earned testing badges saw an increase in their bug fixing commits but output fewer commits and pull requests. 
Finally, 74\% of these survey participants found badges to be useful for learning new practices and were motivated by badges which demonstrate the prospect of reducing manual overhead and standardizing processes across teams and projects.